\documentclass[fleqn,12pt,twoside]{article}
\usepackage{espcrc1}

\usepackage{graphicx}
\usepackage[figuresright]{rotating}
\newcommand{\AmS}{{\protect\the\textfont2
  A\kern-.1667em\lower.5ex\hbox{M}\kern-.125emS}}
\usepackage{amsmath}
\usepackage{amssymb}

\newcommand\e{{\rm e}}
\newcommand\fP{{\mathbb{P}}} 

\title{Polarized Deep Inelastic Diffractive Scattering near the
       Light Cone}
\author{J. Bl\"umlein
\address[DESY]{Deutsches Elektronen--Synchrotron, DESY, Platanenallee 6,
D--15738 Zeuthen, Germany}, and
D. Robaschik\address{BTU Cottbus, Fakult\"at 1, PF 101344, D-03013 
Cottbus, 
Germany}} 
\begin{document}

\maketitle

\begin{abstract}
\noindent
Polarized inclusive deep-inelastic scattering is formulated in the light 
cone expansion. The QCD evolution of the leading twist distribution 
functions is derived. It is shown that the twist--2 contribution 
to the structure functions $g_2^{D(3)}$ is obtained via $g_1^{D(3)}$
by a Wandzura--Wilczek relation.
\end{abstract}
\section{Introduction}
\noindent
Inclusive unpolarized and polarized deeply inelastic diffractive 
scattering at high energies and momentum transfer is one of the important 
processes in lepton--nucleon scattering. As found by experiment, cf.
\cite{DI1}, there are interesting 
relations between the cross sections of these processes and those of
inclusive deeply inelastic scattering:~{\it i)} the scaling violations of 
both processes are quite similar and~{\it ii)} the ratio of the 
differential cross sections in $x$ and $Q^2$ are widely constant in the
whole kinematic domain and are of $O(1/8...1/10)$. Whereas the latter 
aspect cannot be understood with perturbative methods the former calls 
for a rigorous analysis in perturbative QCD. In recent 
analyses~\cite{BR1,BR2} this aspect has been investigated both for the 
unpolarized and the polarized case on the basis of the light--cone 
expansion. By this method the semi-exclusive processes of diffractive 
scattering could be related to forward scattering processes at short 
distances, for which similar evolution equations as in the deep inelastic 
case apply. Moreover a Callan--Gross and Wandzura-Wilczek relation between 
the twist--2 contributions of the diffractive structure functions were 
derived. In this note we give a summary of these papers.

\section{Scattering Cross Section and Formalism}
\noindent
The process of deep inelastic diffractive scattering is $l + p
\rightarrow
l' + N' + hadrons$, with a significant rapidity gap between $N'$ and the
remaining hadrons.
The differential scattering cross section for single--photon
exchange is given by
\begin{equation}
\label{eqD1}
d^5 \sigma_{\rm diffr}
= \frac{1}{2(s-M^2)} \frac{1}{4} dPS^{(3)} 
\frac{e^4}{Q^2} L_{\mu\nu} W^{\mu\nu}~,
\end{equation}
with $L_{\mu\nu}$ and $W_{\mu\nu}$  the leptonic and hadronic tensors. 
Using current conservation, P and T invariance and the hermiticity 
relation for the hadronic tensor one finds a representation of the 
hadronic tensor in terms of four unpolarized and eight polarized structure 
functions~\cite{BR1,BR2}. We will henceforth consider the case of 
small values of $(p_2 - p_1)^2 = t$. In this limit the outgoing and 
incoming proton momenta are related by $p_2 = (1-x_{\fP}) p_1$ and the 
cross 
section depends on two unpolarized and two polarized structure functions 
only
\begin{eqnarray}
\label{eqhadr}
W_{\mu\nu} &=&  \left(g_{\mu\nu} + \frac{q_\mu q_\nu}{q^2}\right) W_1 +
\left(p_{1\mu} - q_\mu \frac{p_1.q}{q^2}\right)\left(p_{1\nu} - q_\nu
\frac{p_1.q}{q^2}\right) \frac{W_2}{M^2}
\nonumber\\ & & +
i \varepsilon_{\mu \nu \lambda \sigma} \frac{q^\lambda S^\sigma}{p_1.q} 
g_1 
+ i \varepsilon_{\mu \nu \lambda \sigma} \frac{q^\lambda(p_1.q S^\sigma 
-S.q p_1^\sigma)}{(p_1.q)^2} g_2~,
\end{eqnarray}
with $M W_1 \rightarrow F_1$ and $ (p_1.q/M) W_2 \rightarrow F_2$ for 
$Q^2, p_1.q \rightarrow \infty$. Eq.~(\ref{eqhadr}) is considered in the
{\sf generalized Bjorken limit:~~} $Q^2 ,p_1.q, p_2.q \rightarrow \infty$ 
and
$x = Q^2/(2p_1.q), \eta = q.(p_2 -p_1)/q.(p_2+p_1)$ = fixed. The 
non-forward  variable $\eta$ is related to another variable often used, 
$x_{\fP}$, by $x_{\fP} = -2 \eta/(1-\eta)$. In the limit $t \rightarrow 0$ 
the 
above structure functions depend on the three variables $x, x_{\fP}$ and
$Q^2$.

Since for diffractive processes the outgoing proton is {\sf well separated 
} in rapidity from the diffractively produced hadrons (rapidity gap), one 
may apply A. Mueller's generalized optical theorem \cite{AM} to calculate 
the scattering cross section. This is done moving the outgoing proton into 
an incoming anti-proton and considering the absorptive part of deep 
inelastic forward scattering off the state $\langle p_1, -p_2, S_1|$
summing over all final-state spins. Note that under this interchange $t$ 
is kept space--like. Due to this operation we may now evaluate the 
Compton--operator  
\begin{eqnarray}
\widehat{T}_{\mu\nu}(x) &=&
RT \left[J_\mu\left(\frac{x}{2}\right)J_\nu\left(-\frac{x}{2}\right) S   
\right] \\
&=&
 -e^2 \frac{\tilde x^\lambda}{2 \pi^2 (x^2-i\epsilon)^2}
 RT
 \left[
\overline{\psi}
\left(\frac{\tilde x}{2}\right)
\gamma^\mu \gamma^\lambda \gamma^\nu \psi
\left(-\frac{\tilde x}{2}\right)
- \overline{\psi}
\left(-\frac{\tilde x}{2}\right)
\gamma^\mu \gamma^\lambda \gamma^\nu \psi
\left(\frac{\tilde x}{2}\right)
\right] S \nonumber
\end{eqnarray}
between the above states for forward scattering. We represent this 
operator in terms of a vector and an axial-vector operator, which are in 
turn related to the associated scalar and pseudo-scalar operators, through 
which we introduce the respective operator expectation values, see 
\cite{BR1,BR2} defining non--forward parton densities 
$f^q_{s,5}(z_+,z_-)$, 
\begin{eqnarray}
\label{eqSCAM}
\langle p_1,-p_2|O^q(\kappa_+ x,\kappa_- x)|p_1,-p_2
\rangle &=& xp_- 
\int Dz ~\e^{-i \kappa_- x p_z} {f}^q_s(z_+,z_-)
,\nonumber\\
\langle p_1,S_1,-p_2|O^q_5(\kappa_+ x,\kappa_- x)|p_1,S_1,-p_2
\rangle &=&
xS
\!\!\!         \int \!\!\!
Dz ~\e^{-i \kappa_- x p_z} {f}^q_5(z_+,z_-)~,
\end{eqnarray}
with $S \equiv S_1$ and $p_{\pm} = p_2 \pm p_1$. Here we neglect 
sub-leading 
components $\propto \pi_- p_+-p_-/\eta$. After passing a series of steps,
see \cite{BR1,BR2}, we may express the hadronic tensor in this 
approximation by one unpolarized and one polarized distribution function,
$f^D$ and $f^D_5$, respectively. For quarks and anti-quarks these 
distribution functions, which are the diffractive parton distributions,
read
\begin{eqnarray}
\label{eqREL}   
f^D_{(5)}(\pm 2\beta,\eta,Q^2)
 =  a
\int_{-\frac{x_{\fP} \pm 2x}{2-x_{\fP}}}
                              ^{-\frac{x_{\fP} \mp 2x}{2-x_{\fP}}} d \rho
f_{(5)}(\rho,\pm 2\beta + \rho(2-x_{\fP})/x_{\fP};Q^2)~. \end{eqnarray}
The upper sign refers to quarks, the lower to anti-quarks, and $a = -1$ in
the unpolarized case, $a=1/x_{\fP}$ in the polarized case, where
$\beta = x/x_{\fP}$.  
\section{Relations between Structure Functions}
\noindent The diffractive structure functions $F_1^D$ and $g_1^D$ obey the
representation
\begin{eqnarray} F_1^D(\beta,\eta,Q^2) &=& \sum_{q=1}^{N_f} e_q^2
\left[
f_q^D(\beta,x_{\fP},Q^2)+\overline{f}^D_q(\beta,x_{\fP},Q^2)\right]\nonumber\\
g_1^D(\beta,\eta,Q^2) &=& \sum_{q=1}^{N_f} e_q^2 \left[
f_{q5}^D(\beta,x_{\fP},Q^2)+\overline{f}^D_{q5}
(\beta,x_{\fP},Q^2)\right]~.
\end{eqnarray}
After some calculation one finds for the twist--2 contributions to the 
hadronic tensor the relations
\begin{eqnarray} 
F_2^D(\beta,\eta,Q^2) &=& 2x  F_1^D(\beta,\eta,Q^2) \nonumber\\  
g_2^D(\beta,\eta,Q^2) &=&   -g_1^D(\beta,\eta,Q^2) + \int_\beta^1
\frac{d \beta'}{\beta'} g_1^D(\beta',\eta,Q^2)~. 
\end{eqnarray}
The Callan--Gross relation between the structure functions depending on 
$\beta$ is modified due to the emergence of $x$, while the 
Wandzura--Wilczek relation holds in the new variable 
$\beta~\epsilon~[0,1]$. The emergence of the integral term in one
of the above relations is due to a basic connection between a 
vector--valued
non--forward distribution function and the associated scalar 
one~\cite{BGR}. The corresponding term  exceptionally cancels in the
Callan--Gross relation but is present in most relations of this type,
see also~\cite{BK,BT}.
\section{Evolution Equations}
\noindent
The evolution equations of the diffractive parton densities can be
formulated starting with the evolution equations for the scalar quark
and gluon operators in the flavor non--singlet and singlet case, see e.g.
\cite{BGR}.
\begin{eqnarray} 
\mu^2 \frac{d}{d \mu^2} O^A(\kappa_+ \tilde{x}, \kappa_- \tilde{x};
\mu^2) = \int D \kappa' \gamma^{AB}(\kappa_+,\kappa_-,\kappa_+',
\kappa_-';\mu^2) 
O_B(\kappa_+' \tilde{x}, \kappa_-' \tilde{x};\mu^2)~,
\end{eqnarray}
with $\mu$ the factorization scale. Forming expectation values as in the
foregoing section one notices that the evolution does not depend
on the value of the light-cone mark $\kappa_+$, which can be set to 0.
Moreover the all-order rescaling relation
\begin{eqnarray} 
\gamma^{AB}(\kappa_+,\kappa_-,\kappa_+',\kappa_-';\mu^2) 
= \sigma^{d_{AB}}
\gamma^{AB}(\sigma\kappa_+,\sigma\kappa_-,\sigma\kappa_+',\kappa_-'; 
\mu^2)~,
\end{eqnarray}
where $d_{AB} = 2+ d_A - d_B, d_q = 1, d_G = 2$, is applied. After some
calculation one finds the following evolution equations
\begin{eqnarray} 
\mu^2 \frac{d}{d \mu^2} f_A^D(\beta,x_{\fP};\mu^2) = \int_\beta^1 \frac{
d\beta'}{\beta'} P_A^B\left(\frac{\beta}{\beta'};\mu^2\right) 
f_B^D(\beta',x_{\fP};\mu^2)~.
\end{eqnarray}
These equations apply both to the unpolarized and polarized diffractive
parton densities of twist--2 to all orders in the coupling constant.
In the calculation the absorptive part of the distribution was taken, 
which identifies the original momentum fraction, cf.~\cite{BR1,BR2},
$\vartheta = z_- + z_+/\eta$
with $2 \beta$ and results in evolution equations with a support  [0,1],
unlike those in $\vartheta$. If compared to the case of deep-inelastic
scattering the evolution is here not in $x$ but in the new variable 
$\beta$. Otherwise the {\sf same} evolution equations are obtained.
This holds also for higher twist operators, for which the argument is 
exactly the same as given in this section, see~\cite{BR1,BR2}.
\section{Conclusions}
\noindent
We derived the twist--2 evolution equations for the unpolarized and
polarized twist--2 diffractive parton distributions considering these
processes in the light--cone expansion at short distances. We also derived
relations between the diffractive structure functions in the unpolarized
and polarized case. The observed similarity of the scaling violations
between deep-inelastic diffractive and deep--inelastic 
structure functions
was shown in the present approach being due to the same structure of
evolution equations which act in the former case on the variable $\beta$
and in the latter on $x$. Polarized deep--inelastic scattering was
not yet observed nor studied in detail in a larger kinematic domain.
It would be interesting to know if the pattern of relations as observed
for unpolarized scattering repeats and the predictions of the present
paper can be verified. The COMPASS experiment and polarized experiments
at future high-energy facilities \cite{FUT} could clarify this.

\end{document}